\documentclass[%
 aip,
 amsmath,amssymb,
 reprint,%
]{revtex4-1}

\usepackage{graphicx}
\usepackage{dcolumn}
\usepackage{bm}

\usepackage[utf8]{inputenc}
\usepackage[T1]{fontenc}
\usepackage{mathptmx}
\usepackage{etoolbox}

\usepackage{bm}
\usepackage{amsthm} 

\usepackage[english]{babel}
\makeatletter
\def\@email#1#2{%
 \endgroup
 \patchcmd{\titleblock@produce}
  {\frontmatter@RRAPformat}
  {\frontmatter@RRAPformat{\produce@RRAP{*#1\href{mailto:#2}{#2}}}\frontmatter@RRAPformat}
  {}{}
}%
\makeatother
\begin{document}

\preprint{AIP/123-QED}

\title[Affine Gauge Theory: A Diffeomorphism Invariant Gauge Theory of Gravity]{Affine Gauge Theory:\\A Diffeomorphism Invariant Gauge Theory of Gravity}
\author{K. Tjandra}
 \email{ktj@nus.edu.sg}
\affiliation{National University of Singapore
}%
\author{K. Singh}%
\affiliation{National University of Singapore
}%

\date{\today}

\begin{abstract}
This paper is a comprehensive investigation of the Affine Gauge Theory (AGT) as a gauge theory of gravity having the same mathematical structure as gauge theories of the other fundamental forces of nature. This mathematical structure consists of a principal fiber bundle over the spacetime manifold that is endowed with an affine connection. The relationship between AGT and various formulations of teleparallel theories of gravity, which are alternatives to General Relativity, are examined. Here, it is argued that the Affine Bundle is the most natural principal fibre bundle for a gauge theory of gravity. AGT is also shown to be strictly diffeomorphism invariant. In particular, an explicit proof of diffeomorphism invariance in AGT is given - showing that AGT possesses the important symmetry of General Relativity, as would be expected from a theory of gravity. Lastly, the claim that AGT is background independent, as General Relativity is, from varying degrees of strictness in the definition of background independence is closely examined. Reasons for why a background independent theory is preferred are also discussed.
\end{abstract}

\maketitle

\begin{quotation}

\end{quotation}

\section{\label{sec:intro}Introduction}
The vision of a Theory of Everything, one theory that would encompass all of Physical reality, remains as one of the grand aspirations of theoretical Physics. The historical trajectory towards unification, from Maxwell’s Electromagnetism in 1873, Electroweak Theory, to the Standard Model in 1967, suggests that perhaps a Theory of Everything is not beyond the Physicist’ reach. Since three of the four fundamental forces have been united in the Standard Model, one may even be tempted to claim that Physics is three fourths along the way towards a Theory of Everything. Gravity, however, proves much less agreeable towards unification. General Relativity, \textit{the} theory of gravity with the metric and curvature as the dynamical variables and so-called Background Independence, differ fundamentally from our Gauge Theoretic description of the other forces, Background Dependent theories with Gauge Potentials as their dynamical variables, which then could be canonically quantized. 

The subject of this paper, Affine Gauge Theory (AGT), is an attempt to produce a theory of gravity that hopefully would be more agreeable towards unification due to its similarity to the gauge theories of the other fundamental forces. AGT, at the same time, also preserves a fundamental characteristic of General Relativity, Diffeomorphism Invariance. AGT is a physical theory in which the mathematical framework of the theory is that of a gauge theory, consisting of a Principal Fibre Bundle and a Principal Connection on that Fibre Bundle, with Affine Group as the Gauge Group, or typical fibre of the Principal Bundle. In calling the subject matter of this thesis Affine Gauge Theory, an emphasis is put on the mathematical structure of AGT, which closely resembles that of the other Gauge Theories.

In order to establish AGT as a valid gauge theory of gravity, we shall proceed cautiously starting out with laying the foundations and reviewing other attempts of presenting a gauge theory of gravity. 
In section \ref{chap:connectiondebate}, a debate from two camps within the literature regarding how to best formulate Teleparallel Equivalent of General Relativity (TEGR) as a gauge theory is summarized. This thesis argues that both sides of that debate do not give a satisfactory account of TEGR as a gauge theory. section \ref{chap:ans} presents a third alternative, the Affine Gauge Theory. Affine Bundle is argued to be the most natural principal fibre bundle for a gauge theory of gravity. Moreover, relationship between AGT and TEGR is rigourously discussed. sections \ref{chap:transinv}, \ref{chap:candifinv}, and \ref{chap:condifinv} shows mathematical proofs of the Translational Gauge Invariance of AGT, Diffeomorphism Invariance of the canonical 1-form, and the Diffeomorphism Invariance of the flatness of the Frame Bundle Connection respectively. These proofs then build up to section \ref{chap:difinvtheories} where the Diffeomorphism Invariance of both TEGR and AGT is shown and section \ref{chap:relationship} where the relationship between Translational Gauge Invariance and Diffeomorphism Invariance of AGT is firmly established. 

\section{The Connection Debate\label{chap:connectiondebate}}

In \cite{Pereira1Aldrovandi}, Aldrovandi and Pereira presented a version of teleparallel theory of gravity called the Teleparallel Equivalent of General Relativity (TEGR). Teleparallel theory of gravity itself was first introduced by Einstein in 1928. Teleparallel means distant parallelism, which is realized through a flat connection, or a connection with zero curvature. In place of curvature, which is the dynamical variable in GR, torsion takes the role of the dynamical variable in TEGR. Torsion is a function of the tetrad, $B$ (or tetrad component according to Aldrovandi et al.) and the flat Lorentz or Spin Connection, $A$. Equation (4.52) of \cite{Pereira1Aldrovandi} defines (local) torsion as follows, 

\begin{equation}
    T^a_{\mu\nu} = \partial_\mu B_\nu^a - \partial_\nu B_\mu^a + A^a_{b\mu}B^b_\nu - A^a_{b\nu} B^b_\mu \label{eq:per1}
\end{equation}
The equation above is written in tensorial formalism, as is the case for most of the works of Pereira et al., in \cite{Pereira1Aldrovandi}, \cite{Pereira2}, \cite{Pereira3Obukhov}. In the language of differential geometry which have been used in the thesis thus far, which would make such equations simpler and mathematical relationships more apparent, equation \ref{eq:per1} can be written as

\begin{equation}
    T = dB + A \wedge B \label{eq:tbab}
\end{equation}
which is equivalent to the definition the local representation, $T$, of Torsion, $\Theta$. This equivalence is established by the identification of Aldrovandi's tetrad component and pure Lorentz connection, with our local representation of the canonical 1-form and our local representation of the Frame Bundle connection respectively. 

Utilizing their definition of torsion, Aldrovandi and Pereira constructed a theory which is shown to be able to produce Lagrangians which couple mass to torsion and which are equivalent to Lagrangians of GR that couple mass to curvature. Thereby, Aldrovandi establishes the "Equivalent of GR" part of TEGR. 

As such, in TEGR, we have a theory in which gravity is mediated through torsion. However, the claims of \cite{Pereira1Aldrovandi} goes beyond this. Aldrovandi argues that this TEGR is a gauge theory. In fact, the gauge theoretic nature of the theory is the motivating thrust behind TEGR. The objective of Aldrovandi's presentation of TEGR is to present a gauge theory of gravity, not just any theory of gravity. 

It is claimed that TEGR is a gauge theory of translations. Aldrovandi pointed out that the (component of the) tetrad $B$ is a 1-form valued in the lie-algebra of the translations group, $T^4$ or $\mathbb{R}^4$,  and as such, can be taken to be a connection on the bundle of translations. In this interpretation, TEGR is then a theory whose dynamical variable, $B$, is a gauge potential, or the local representation of a connection on a principal bundle. In short, Aldrovandi argues that TEGR is a \textit{bona fide} gauge theory of translations. 

A sharp criticism to the interpretation put forward in \cite{Pereira1Aldrovandi} was produced by Fontanini in \cite{Fontanini1}.  Fontanini argues that the interpretation that considers $T$ a gauge field strength of the translational gauge field $B$, is erroneous. He rightly pointed out that if we have a principal bundle of translations, with connection $\omega_T$, the curvature of that connection, $\Omega_T$, or locally the gauge field strength, will be given by 

\begin{equation}
    \Omega_T = d\omega_T + \omega_T \wedge \omega_T
\end{equation}
or locally 

\begin{equation}
    F_T = dB + B \wedge B
\end{equation}
and not equation \ref{eq:tbab}. 
Fontanini rightly pointed out that the heuristic (or rather, constructivist) introduction of $A$ in

 \begin{equation}
     T = dB + A \wedge B
 \end{equation}
 
implies an implicit gauging of the Lorentz group. Therefore, TEGR is not simply a gauge theory of translations, rather, it is a theory of gravity based on the Frame Bundle, $LM$, with a Frame Bundle connection, $\omega$, locally represented by $A$, and a canonical 1-form, $\theta$, locally represented by $B$. 

Fontanini then put forward an alternative, a radically different approach to TEGR in \cite{Fontanini1} which was clarified and further expounded in \cite{Fontanini3Huguet}. This new approach is called the Cartan approach based on the purely mathematical works of Sharpe in \cite{Sharpe}. In this book, Sharpe sought to reconstruct differential geometry based on the so-called Cartan connection instead of the conventional Ehresman connection, or simply the connection as have been used in this thesis thus far. 

Fontanini's Cartan approach to TEGR is defined on the Frame Bundle, $LM$, with a Cartan connection, $\omega_c$. This Cartan connection is a 1-form defined on the $LM$ and valued in the lie-algebra of the Affine Group, $\mathfrak{a}(n,\mathbb{R})$, not just in the lie algebra of the General Linear Group, $\mathfrak{gl}(n,\mathbb{R})$, i.e.  

\begin{equation}
    \omega_c : T_pLM \rightarrow \mathfrak{a}(n,\mathbb{R}) 
\end{equation}

This Cartan connection then takes the place of translational connection in Aldrovandi's version of TEGR. Fontanini showed that the (local) torsion, $T$, can be recovered from some form of curvature from this Cartan connection. Moreover, this Cartan connection, $\omega_c$,  can be split into two components, $\omega_{gl}$ valued in the lie-algebra of the General Linear Group and $\omega_r$ valued in the lie-algebra of the n-dimensional real space, $\mathbb{R}^n$

\begin{equation}
    \omega_c = \omega_{gl} + \omega_r
\end{equation}
These two components are equivalent to the Frame Bundle Connection and the Canonical 1-form both defined in $LM$. 

\begin{equation}
    \omega_c = \omega + \theta \label{eq:omomthe}
\end{equation}

Equation \ref{eq:omomthe} implies that, in terms of our standard and conventional interpretation of differential geometry, the new 1-form $\omega_c$ is not a connection 1-form at all, but the summation of two 1-forms, one of which is the standard Ehresman connection, and the other is the canonical 1-form. Calling this new 1-form a connection is, at the very least, strange. As clearly shown in \cite{Sharpe}, the only part that both determines the horizontality of $T_pP$ of some principal bundle $P$ and defines the covariant derivative is only the first component, which is nothing but the Ehresman connection itself (after all, $\omega_c$ has no kernel). 

Just as Aldrovandi and Pereira sought to interpret their TEGR as a gauge theory by redefining the curvature or gauge field strength, Fontanini sought to interpret his approach to TEGR as a gauge theory by redefining the connection. Fontanini found such a redefinition in the works of Sharpe in \cite{Sharpe}. While the arguments of \cite{Sharpe} which are arguments of pure abstract mathematics which sought to reconstruct differential geometry on a supposedly firmer footing may be valid, these questions do not concern the physicist and is beside the point of our physical scientific endeavor. After all, what we are interested in is a gauge theory of gravity which has the same mathematical structure as the other gauge theories of the other fundamental forces of nature. These other gauge theories consist of a principal bundle, and an Ehresman connection, not a Cartan connection. 

In summary,  \cite{Pereira1Aldrovandi} puts forward TEGR, which is a theory of gravity that can be defined in the frame bundle, $LM$ with a flat Frame Bundle connection, $\omega$, canonical 1-form, $\theta$, and torsion, $T$.  This theory is then given an interpretation as a gauge theory of translations, which implies that the theory is defined on the translations bundle $TrM$ with a translation connection, $\theta$, and a redefined curvature or gauge field strength, $T$. Fontanini on the other hand, proposes an approach to TEGR defined in the frame bundle, $LM$ with a redefined connection, the Cartan connection, $\omega_c$, and a local curvature component, $T$. Both proposals do not meet the objective of putting forward a theory of gravity with the same mathematical structure as the other gauge theories for the other fundamental forces of nature. 

\section{AGT as the Gauge Theoretic Answer\label{chap:ans}}

How can TEGR be formulated as a gauge theory? How can it be defined in a principal bundle, whose connection and curvature realizes local torsion as defined by Aldrovandi and Pereira in \cite{Pereira1Aldrovandi}? The answer lies in the Affine Gauge Theory (AGT). Affine Gauge Theory is a theory defined in the Affine Bundle, with its conventional Ehresman connection. 

Let us recall that in TEGR, we have a flat Frame Bundle connection, 

\begin{equation}
    \omega : T_pLM \rightarrow \mathfrak{gl}
\end{equation}
and the canonical 1-form

\begin{equation}
    \theta: T_pLM \rightarrow \mathbb{R}^n \label{eq:the}
\end{equation}

Since the lie-algebra of the space $\mathbb{R}^n$ treated as a group is also $\mathbb{R}^n$, denoted by $\mathfrak{r}$, \ref{eq:the} can also be written as

\begin{equation}
\theta: T_pLM \rightarrow \mathfrak{r}
\end{equation}

Under a local section, $\sigma$, $\omega$ is represented on the base manifold, $M$, by
\begin{gather}
    A: T_xM \rightarrow \mathfrak{gl} \\
    A = \sigma^*\omega 
\end{gather}
 and $\theta$ is represented on the base manifold, $M$, by 
 \begin{gather}
     B:T_xM \rightarrow \mathfrak{r} \\ 
     B = \sigma^* \theta
 \end{gather}

Furthermore, it has been shown by Fontanini et al. in \cite{Fontanini1} \cite{Fontanini3Huguet} that the novel 1-form $\omega_c$ can realize Torsion as a component of its curvature and that
\begin{equation}
    \omega_c = \omega + \theta
\end{equation}
Under a local section, $\sigma$, the above equation becomes 
\begin{align}
    \sigma^* \omega_c &= \sigma^* \omega + \sigma^* \theta \\
    A_c &= A + B
\end{align}
where $\omega_c$ and $A_c$ are valued in the product space of the vector spaces $\mathfrak{gl}$ and $\mathfrak{r}$, the space of the lie-algebra of the Affine Group, 
\begin{equation}
    \mathfrak{a} = \mathfrak{gl} \times \mathfrak{r}
\end{equation}
As such, we have a 1-form, $A_c$, defined on the base manifold that can realize torsion $T$, which is the dynamical variable in TEGR. In order for this 1-form on the manifold to be considered a proper local representation of a connection, it needs to be a pull-back of a principal bundle connection defined on some principal bundle. To divorce the core of the argument from Fontanini's proposal, we consider a one form that is locally the same as $A_c$ but in fact is a local representation of a principal bundle
\begin{gather}
    \tilde{A}: T_xM \rightarrow \mathfrak{a} \\
    \tilde{A} = A + B \\ 
    \tilde{A} = \tilde{\sigma}^* \tilde{\omega}
\end{gather}
Since the connection $\tilde{\omega}$ is valued in the lie-algebra of the Affine Group, this new principal bundle necessarily needs to have the Affine Group as its typical fibre. Therefore the only principal bundle that produce a standard gauge theory which realizes torsion, as a part of its curvature, is the Affine Bundle. Thus, Affine Bundle is the most natural principal fibre bundle for a gauge theory of gravity. This Affine Gauge Theory answer is briefly alluded to by Fontanini et al. in \cite{Fontanini1} and \cite{Fontanini2Delliou}, and also proposed in Pereira's response to \cite{Fontanini1} in \cite{Pereira2}. 

\subsection{Affine Bundle and Frame Bundle Correspondence\label{Correspondence}}

In the current section, the correspondence between the Affine Bundle and the Frame Bundle will be studied in a rigorous manner. This correspondence is the mathematical foundation that renders AGT equivalent to TEGR in the regime that is of interest to us. This section follows Kobayashi's book, \cite{Kobayashi} section III.3, closely.

The Affine Bundle and the Frame Bundle are related through the maps $\beta: A(M) \rightarrow L(M)$ and $\gamma: L(M) \rightarrow A(M)$, defined as 
\begin{align}
    &\beta((p;E_1,E_2, ..., E_n))=(E_1,E_2, ..., E_n) \\
    &\gamma((E_1,E_2, ..., E_n))=(O_E;E_1,E_2, ..., E_n)
\end{align}
where $u=(E_1,E_2, ..., E_n)$ are frames at some spacetime point $x \in M$, i.e. $u \in Gl(n,\mathbb{R})$ and $\tilde{u}=(O_E;E_1,E_2, ..., E_n)$ are frames together with a choice of origin at some spacetime point $x \in M$, i.e. $\tilde{u} \in A(n,\mathbb{R})$. $Gl(n,\mathbb{R})$ and $A(n,\mathbb{R})$ are the typical fibres of the principal bundles the Frame Bundle, $LM$, and the Affine Bundle, $AM$, respectively. Note that while $\gamma$ is injective, $\beta$ is surjective. Furthermore, $\gamma$ is a choice of origin in the Affine Bundle and thus defines a privileged section. This privileged section can be written as 
\begin{equation}
    \tilde{\sigma} = \gamma \cdot \sigma \label{eq:priv}
\end{equation}

Thus, it is manifest that given a section in $L(M)$, $\sigma:M \rightarrow L(M)$, there is a section in $A(M)$, $\tilde{\sigma}:M\rightarrow A(M)$.

Moreover, the $\gamma$ map, relates the two components of the Affine Connection, $\tilde{\omega}_L$ and $\tilde{\omega}_R$, valued in the lie algebra of $Gl(n,\mathbb{R})$ and $\mathbb{R}^n$ respectively, to the Frame Bundle Connection, $\omega$, and the Canonical 1-form, $\theta$, on the Frame Bundle

\begin{align}
    \tilde{\omega} &= \tilde{\omega}_L + \tilde{\omega}_R \\
    \gamma^* \tilde{\omega}_L &= \omega \\
    \gamma^* \tilde{\omega}_R &= \theta
\end{align}

Hence the Affine Connection, $\tilde{\omega}$, on the Affine Bundle, is related to the Frame Bundle Connection, $\omega$, and the Canonical 1-form, $\theta$, on the Frame Bundle in the following way,
\begin{equation}
    \gamma^* \tilde{\omega} = \omega + \theta \label{eq:reltilnotil}
\end{equation}

The relationship of the local representations of these 1-forms are even closer since from the equations above we get
\begin{align}
    \widetilde{A} &= \tilde{\sigma}^*\tilde{\omega} \\
    &= (\gamma \cdot \sigma)^* \tilde{\omega} \\
    &= \sigma^* \gamma^* \tilde{\omega} \\
    &= \sigma^* (\omega + \theta) \\
    \widetilde{A}&= \sigma^* \omega + \sigma^* \theta \label{eq:loctild}
\end{align}
I.e. the local representation of the Affine Connection, $\widetilde{A}$, is nothing but the summation of the local representations of the Frame Bundle Connection and the Canonical 1-form. We can write equation \eqref{eq:loctild} as 
\begin{align}
    \widetilde{A} &= A + B \label{eq:ab} \\
    \widetilde{A} &= A_b^a \Phi_a^b + B^a P_a
\end{align}
since 
\begin{align}
    \sigma^* \omega &= A_b^a \Phi_a^b \\
    \sigma^* \theta &= B^a P_a
\end{align}
where $\Phi_a^b$ and $P_a$ are the generators of the lie algebras of groups $Gl(n,\mathbb{R})$ and $\mathbb{R}^n$ respectively. Recall how these two 1-forms on the base manifold, $A$ and $B$, have been seen before in our examination of Aldrovandi and Pereira's TEGR. Physics is done in the local representations of the global mathematical formulation. Therefore, the physics that can be produced from the Affine Connection is exactly the physics of TEGR.

Furthermore, the curvature of the Affine Bundle, under local representation, is given by 
\begin{equation}
    \tilde{F}_b^a = d\tilde{A}^a + \tilde{A}_c^a \wedge \tilde{A}_b^c \label{eq:fftilde}
\end{equation}
Expanding equation \eqref{eq:fftilde} in terms of $A$ and $B$, using equation \eqref{eq:ab} gives us
\begin{equation}
    \tilde{F}_b^a = dA_b^a + A_c^a \wedge A_b^c + dB^a + A_b^a \wedge B^b
\end{equation}
which can be simplified in terms of curvature and torsion in the Frame Bundle, which gives us 
\begin{equation}
    \tilde{F}_b^a = F_b^a + T_b^a
\end{equation}

\subsection{AGT and TEGR Correspondence\label{sec:atcorr}}

In the previous section the correspondence between the Affine Bundle and the Frame Bundle has been established. The correspondence between the 1-forms defined on the two bundles has also been examined. These 1-forms include the Frame Bundle Connection 1-form and the Canonical 1-form on the Frame Bundle, the building blocks of TEGR, and the Affine Bundle Connection 1-form, the building block of AGT. How exactly then does this two theories correspond to one another? Beside the 1-forms, we need to examine the teleparallelism condition of TEGR, and what this teleparallelism in the Frame Bundle look like for AGT.

For the Frame Bundle Connection, to render the theory Teleparallel in the Frame Bundle, a connection with 
\begin{equation}
    A = \phi^{-1} d\phi \label{eq:phidphi}
\end{equation}
is chosen as the flat connection. Where $\phi \in Gl(n,\mathbb{R})$.

To show its flatness, we compute the curvature of the above connection. 
\begin{align}
    F_b^a &= dA_b^a+A_c^a \wedge A_b^c \\
    &= d[\phi^{-1} d\phi]_b^a + [\phi^{-1}d\phi]_c^a \wedge [\phi^{-1}d\phi]_b^c \\
    &= [d\phi^{-1}]_c^a \wedge [d\phi]_b^c + [\phi^{-1}]_d^a [d\phi]_c^d \wedge [\phi^{-1}]_e^c [d\phi]_b^e \label{eq:f}
\end{align}
Evaluating the second term
\begin{align}
    [\phi^{-1}]_d^a [d\phi]_c^d \wedge [\phi^{-1}]_e^c [d\phi]_b^e &= [\phi^{-1}]_d^a [d\phi]_c^d [\phi^{-1}]_e^c \wedge  [d\phi]_b^e\\
    &= [\phi^{-1}]_d^a \{ d([\phi]_c^d[\phi^{-1}]_e^c) - \\ &[\phi]_c^d [d\phi^{-1}]_e^c \} \wedge [d\phi]_b^e \\
    &= [\phi^{-1}]_d^a \{ d(\mathbb{1}_e^d) -\\ &[\phi]_c^d [d\phi^{-1}]_e^c \} \wedge [d\phi]_b^e \\
    &= - [\phi^{-1}]_d^a  [\phi]_c^d [d\phi^{-1}]_e^c  \wedge [d\phi]_b^e \\
    &= - \delta_c^a [d\phi^{-1}]_e^c  \wedge [d\phi]_b^e \\
    &= - [d\phi^{-1}]_e^a  \wedge [d\phi]_b^e \label{eq:f2}
\end{align}
Substituting equation \eqref{eq:f2} into \eqref{eq:f} gives us
\begin{align}
    F_b^a &= [d\phi^{-1}]_c^a \wedge [d\phi]_b^c - [d\phi^{-1}]_e^a  \wedge [d\phi]_b^e \\
    F_b^a &= 0 \qed \label{eq:flat}
\end{align}

Showing that the connection $A=\phi^{-1} d\phi$ is a flat connection. Thus, such connections can be used in TEGR as a connection that realizes distant parallelism in the Frame Bundle. 

The corresponding component of the Affine Connection in the Affine Bundle can also be represented locally as $A=\phi^{-1}d\phi$, since as we recall in equation \eqref{eq:loctild}, the local representation of the Affine Connection is the summation of the local representation of the Frame Bundle Connection and the Canonical 1-form. As such, in the Affine Bundle corresponding to a flat Frame Bundle, equation \eqref{eq:ab} becomes
\begin{equation}
    \tilde{A} = \phi^{-1} d\phi + B 
\end{equation}
With the $Gl(n,\mathbb{R})$ component of the local Affine Connection specified, the curvature of the Affine Connection can be computed. On the Affine Bundle, the curvature of the Affine Connection is given by 
\begin{equation}
    \tilde{\Omega} = D_{\tilde{\omega}}\tilde{\omega}
\end{equation}
or locally
\begin{equation}
    \tilde{F} = d\tilde{A}+ \tilde{A} \wedge \tilde{A}
\end{equation}
expanding this in terms of $A$ and $B$ gives us
\begin{align}
    \tilde{F} &= d(A+B) + (A+B) \wedge (A+B) \\
    \tilde{F} &= dA + dB + A \wedge A + A \wedge B + B \wedge A + B \wedge B \label{eq:abelb}
\end{align}
gathering the terms that give us $F$ and $T$, and since the generators of the lie-algebra of $\mathbb{R}^n$ is commutative or in other words, translations are Abelian, equation \eqref{eq:abelb} becomes
\begin{align}
\tilde{F} &= dA + A \wedge A + dB + A \wedge B + B \wedge A \\
\tilde{F} &= F + T + B \wedge A
\end{align}
Expressing the last term in matrix representation gives us
\begin{equation}
    B \wedge A = \begin{pmatrix}
        0 & B \\ 0 & 0
    \end{pmatrix} \wedge \begin{pmatrix}
        A & 0 \\ 0 & 0
    \end{pmatrix}
\end{equation}
which clearly shows that the last term is equals to zero. Therefore, 
\begin{equation}
    \tilde{F} = F + T
\end{equation}
The curvature of the Affine Connection is the summation of the curvature of the Frame Bundle connection and the Torsion. Now, since the Frame Bundle connection that has been chosen for TEGR is the flat connection, $A=\phi^{-1} d\phi$, $F=0$, the curvature of the corresponding Affine connection on the Affine Bundle reduces to 
\begin{equation}
    \tilde{F}=T \label{eq:ft}
\end{equation}
i.e. the local torsion $T$ defined on the Frame Bundle. As such, in AGT, in the regime that corresponds to TEGR, the curvature of the Affine connection is purely the component which corresponds to Torsion. Locally, in AGT, the gauge field strength, which is defined as the local curvature is equal to the local Torsion as defined on the Frame Bundle. As such, in AGT, we have a true gauge theory which realizes the Torsion of TEGR, which then can be used to construct Lagrangians that are equivalent to the Lagrangians of GR via the Equivalence of GR approach of TEGR. In short, AGT is a gauge theory of gravity.  

It is important to note however, that the curvature of the Affine Connection is not zero. After all, this curvature is, locally, our gauge field strength. If it is zero, there would not be any non-trivial dynamics. As such, AGT cannot be described to be teleparallel. Which is why this thesis have chosen to call the theory AGT and not something along the lines of the Affine approach to TEGR. AGT, so far, have also been discussed as such, not as an approach of TEGR, but as a different theory that attributes gravity to torsion rather than curvature in the spirit of TEGR. More importantly, it is a \textit{bona fide} gauge theory. 

\section{Translational Gauge Invariance \label{chap:transinv}}

In the previous section, the argument that puts AGT forward as the true gauge theory of gravity has been clearly laid out. Since the theory is a gauge theory, the theory necessarily has gauge invariance. The physics that can be derived from such a theory is independent of gauge transformations or change in the local section of the bundle. However, does the theory retain its form and equivalence to TEGR given a purely translational gauge invariance? This question will be examined in the current section.

A gauge transformation in fibre bundle formulation is precisely a vertical automorphism of the principal bundle. A vertical automorphism, $\mathcal{F}$, is a map from $A(M)$ to $A(M)$ that maps every $\tilde{u} \in A(M)$
\begin{equation}
    \mathcal{F}: \tilde{u} \rightarrow \mathcal{F}(\tilde{u})  
\end{equation}
such that $\tilde{u}$ and $\mathcal{F}(\tilde{u})$ both belong to the same fibre, rendering the map vertical,
\begin{equation}
    \tilde{\pi}(\mathcal{F}(\tilde{u})) = \tilde{\pi}(\tilde{u})
\end{equation}

Furthermore, the map also fulfils the following condition 
\begin{equation}
    \mathcal{F}(\tilde{u}\tilde{g}) = \mathcal{F}(\tilde{u}) \cdot \tilde{g}
\end{equation}

In particular, the kind of vertical automorphism that is of interest in this section are those related to the translations which form a subgroup of the Affine Group. Under this translational vertical automorphism, every $\tilde{u}$ is mapped into 
\begin{equation}
    \mathcal{F}(\tilde{u})=\tilde{u}\phi(\tilde{u})
\end{equation}
where 
\begin{align}
    \phi: A(M) \rightarrow A(n,\mathbb{R}) \\
    \phi (\tilde{u}\tilde{g}) = \tilde{g}^{-1} \phi (\tilde{u}) \tilde{g}
\end{align}
and more specifically, $\phi(\tilde{u})$ and $\tilde{g}$ is always an element of the translational subgroup of $A(n,\mathbb{R})$ for all $\tilde{u} \in A(M)$.

Under such an automorphism, the local representations of the Affine Connection, $\tilde{A}$, transforms into 
\begin{align}
    \tilde{A}' &= \tau \tilde{A} \tau^{-1} + \tau d\tau^{-1} \\
    &= \tau (A_b^a \Phi_a^b +B^aP_a) \tau^{-1} + \tau d\tau^{-1} \label{eq:aprime}
\end{align}
where $\tau$ can be represented by a matrix of the form 
\begin{equation}
    \tau = \begin{pmatrix}
        \mathbb{1}_{n\times n} & c \\
        0 & 1
    \end{pmatrix}
\end{equation}
Under such a representation, equation \eqref{eq:aprime} can be calculated explicitly which yields
\begin{equation}
    \tilde{A}' = A + (B^a -dc^a - A_b^a c^b) e_a
\end{equation}
the second and third terms in the bracket can be denoted by
\begin{equation}
    D_Ac = (dc^a + A_b^a c^b) e_a
\end{equation}
where $D_A$ denotes the covariant derivative with respect to connection $A$, since these terms are identical to the covariant derivative of $c$, the translation gauges. 

As such, since 
\begin{equation}
    \tilde{A} = A + B
\end{equation}
and similarly, it can be written that
\begin{equation}
    \tilde{A}' = A'+B'
\end{equation}
we conclude that under a translational gauge transformation, the components of the Affine Bundle Connection transforms in the following way
\begin{align}
    A &\rightarrow A' = A \\
    B &\rightarrow B' = B-D_Ac
\end{align}

Moreover, as for the curvature of the Affine Bundle Connection, under a translational gauge transformation, $\tilde{F}$ transforms into 
\begin{align}
    \tilde{F}' &= F + T + D_A(D_A c) \\
    &= F + T - F_b^ac^be_a
\end{align}

In terms of the components of the curvature of the Affine Bundle Connection
\begin{align}
    F &\rightarrow F'=F \\
    T &\rightarrow T'=T-F_b^ac^be_a
\end{align}

Which implies that the component of the curvature of the Affine Bundle Connections that corresponds to Frame Bundle Connection in general is invariant under a translational gauge transformation while the component that corresponds to Torsion in the Frame Bundle is not invariant, in general. Rather, it transforms depending on curvature of the Frame Bundle Connection. However, since as discussed in section \ref{sec:atcorr} the Frame Bundle Connection that is of interest to this study are flat connections, i.e. 
\begin{equation}
    F = 0
\end{equation}
the curvature of the Affine Bundle Connection in the regime that is equivalent to the teleparallel regime in the Frame Bundle, along with its individual components are invariant under a translational gauge. 
\begin{align}
    \tilde{F} &\rightarrow \tilde{F}'=\tilde{F} \label{eq:one}\\
    F &\rightarrow F' = F \label{eq:two}\\
    T &\rightarrow T'=T \label{eq:three}
\end{align}
Which implies an explicit translational gauge invariance, which is characteristic of the AGT. 

\section{Diffeomorphism Invariance of the Canonical 1-form \label{chap:candifinv}}

Having established AGT as a true gauge theory of gravity in section \ref{chap:ans} and its explicit translational gauge invariance in section \ref{chap:transinv}, the question remains as to whether AGT and TEGR possess diffeomorphism invariance. This question is crucial as diffeomorphism invariance and not just translational symmetry is properly understood to be the defining symmetry of GR. Before examining diffeomorphism invariance of AGT and TEGR, the effect of diffeomorphisms on the crucial building block of TEGR, and by extension AGT, the canonical 1-form will be studied in this section.

\subsection{Induced Automorphism of the Frame Bundle}

Suppose a diffeomorphism, $f$, of the base manifold, $M$, is given
\begin{equation}
    f: M \rightarrow M
\end{equation}
A diffeomorphism maps every point in $M$ to another point in $M$, while preserving the differential structure of $M$. As such, the push-forward of a vector on $M$ can be defined,
\begin{equation}
    f_*: T_xM \rightarrow T_{f(x)}M
\end{equation}
Now, since the frames, $u=(E_1,E_2,...,E_n)$, in a frame bundle, $LM$, consist of vectors on M, $E_i \in T_xM$. Every diffeomorphism, $f$, on the base manifold naturally induces a principal bundle automorphism, $\mathcal{F}$, on the frame bundle defined as

\begin{equation}
\label{eq:indfra}
    \mathcal{F}(u) = (f_*E_1, f_*E_2, ..., f_*E_n)
\end{equation}

This induced automorphism will be the main focus of this section. More precisely, the effects of this automorphism on the Canonical 1-form will be studied. 

\subsection{\label{sec:luku}Local Representations of $l_u$ and $k_u$}

Given a specific frame of $M$, $u=(E_1,E_2,...,E_n)$, locally, 
\begin{equation}
    l_u(\xi)=\xi^a E_a^b \frac{\partial}{\partial x^b}
\end{equation}
where $(x^1,x^2,...,x^n)$ are the local coordinates in some chart of $M$.

In these coordinates, $k_u$ becomes
\begin{align}
    k_u(\bar{X}^c \frac{\partial}{\partial x^c}) & = \langle F^a_b dx^b \otimes e_a , \bar{X}^c \frac{\partial}{\partial x^c} \rangle \\
    & = \bar{X}^c \delta_c^b F_b^a e_a \\
    & = \bar{X}^b F_b^a e_a \label{eq:kuloc}
\end{align}
where $F$ is the matrix inverse of $E$ and $e_a$ is a basis vector of $\mathbb{R}^n$. Thus, \eqref{eq:kuloc} can be written as
\begin{equation}
    k_u(\bar{X}^a \frac{\partial}{\partial x^a}) = \bar{X}^b (E^{-1})_b^a e_a
\end{equation}
More specifically
\begin{equation}
\label{eq:kubas0}
    k_u(\frac{\partial}{\partial x^a}) = (E^{-1})_a^b e_b
\end{equation}
or writing $k_u$ in the basis 1-forms
\begin{equation}
\label{eq:kubas}
    k_u= (E^{-1})_a^b dx^a e_b
\end{equation}

This will be useful later when $\mathcal{F}^*\theta_{u'}$ is computed

\subsection{Local Section}

With a local section, $\sigma$, which maps every $x \in U$, where $U$ is a neighborhood of $M$, to a $u \in \pi^{-1}(x)$, we can write every $u \in LM$ as
\begin{equation}
    u = \sigma(x) \cdot g
\end{equation}
where $g \in Gl(n,\mathbb{R})$.

A local section on the frame bundle can be written as 
\begin{equation}
    \sigma(x)=(\frac{\partial}{\partial x^1}, \frac{\partial}{\partial x^2}, ..., \frac{\partial}{\partial x^n})
\end{equation}
Where $(x^1,x^2,...,x^n)$ are the local coordinates in some chart of $M$, as defined earlier in section \ref{sec:luku}.

When the automorphism defined in equation \eqref{eq:indfra} is applied on the local section, 
\begin{equation}
\label{eq:fdel}
    \mathcal{F}(\sigma(x)) = (f_*\frac{\partial}{\partial x^1}, f_*\frac{\partial}{\partial x^2}, ..., f_*\frac{\partial}{\partial x^n})
\end{equation}
In the basis vectors of $T_{x'}M$, the frames in equation \eqref{eq:fdel} becomes
\begin{equation}
\label{eq:phi}
    f_* \frac{\partial}{\partial x^a}\bigg|_x = \Phi_a^b(x)\frac{\partial}{\partial x^b}\bigg|_{x'}
\end{equation}
Note that $\Phi_a^b(x) \in Gl(n,\mathbb{R})$ is specified by $f$, and dependent on $x$. This is how the basis vectors on the base manifold, $\frac{\partial}{\partial x^a}$, transforms under a push-forward along the diffeomorphism. 

On the other hand, the pull-back of the basis 1-forms along the diffeomorphism is given by 
\begin{equation}
\label{eq:phione}
    f^*dx^a |_{x'} = \Phi_b^a dx^b |_x
\end{equation}
since
\begin{align}
    \langle dx^a |_{x'} , f_* \frac{\partial}{\partial x^c} \bigg|_x \rangle &= \Phi_c^b \langle dx^a |_{x'} , \frac{\partial}{\partial x^b} \bigg|_{x'} \rangle \\
    \langle f^* dx^a , \frac{\partial}{\partial x^c} \rangle &= \Phi_c^b \delta_b^a = \Phi_c^a
\end{align}

Now, under the automorphism, $\mathcal{F}$, $u$ is mapped onto
\begin{align}
    u' &= \mathcal{F}(u) \\
    &= \mathcal{F}(\sigma(x)\cdot g) \\
    &= \mathcal{F}(\sigma(x)) \cdot g \label {eq:factg}
\end{align}

Substituting equation \eqref{eq:phi} into \eqref{eq:factg}
\begin{align}
    u'&= \{ g \cdot \Phi_a^b(x) \frac{\partial}{\partial x^b} \}_{a=1, 2, ..., n} \\
    &= \{ \Phi_a^b(x) g_b^c \frac{\partial}{\partial x^c} \}_{a=1, 2, ..., n} \\
    &= \{ (\Phi \cdot g)_a^b \frac{\partial}{\partial x^b} \}_{a=1, 2, ..., n}
\end{align}

With $u'$ now written in the form $\{ (E)_a^b \frac{\partial}{\partial x^b} \}_{a=1, 2, ..., n}$, equation \eqref{eq:kubas} gives us
\begin{equation}
    k_{u'} = ((\Phi \cdot g)^{-1})_a^b dx^a e_b |_{x'}
\end{equation}
or
\begin{equation}
\label{eq:kupri}
    k_{u'} = (\Phi \cdot g)^{-1} \cdot dx^a e_a |_{x'}
\end{equation}

\subsection{Proof of Diffeomorphism Invariance of the Canonical 1-form}

Using the results that were derived in the previous sections, now the diffeomorphism invariance of the canonical 1-form can be shown, i.e.
\begin{equation}
    \mathcal{F}^* \theta = \theta
\end{equation}
or 
\begin{equation}
\label{eq:ditheta}
    \mathcal{F}^* \theta_{u'} = \theta_u
\end{equation}

To show \eqref{eq:ditheta}, first let us compute the pull-back of the Canonical 1-form at $u'$,
\begin{equation}
    \mathcal{F}^*\theta_{u'}(X_u) = \theta_u' (\mathcal{F}_* X_u)
\end{equation}

Using the definition of the Canonical 1-form, we get
\begin{align}
    \mathcal{F}^*\theta_{u'}(X_u) &= k_u' (\pi_*\mathcal{F}_* X_u) \\
    &= k_u' (f_* \pi_* X_u) \label{eq:inskupri}
\end{align}
where $f:M\rightarrow M$ denotes a diffeomorphism on $M$.
Substituting equation  \eqref{eq:kupri} into \eqref{eq:inskupri} gives us 
\begin{align}
    \mathcal{F}^*\theta_{u'}(X_u) &= \langle (\Phi \cdot g)^{-1} \cdot dx^a e_a |_{x'} , f_* \bar{X}_x \rangle \\ 
    &= \langle (\Phi \cdot g)^{-1} \cdot f^* dx^a e_a |_{x'} , \bar{X}_x \rangle \label{eq:brak1}
\end{align}
where $\bar{X}_x = \pi_*X_u$. Substituting equation \eqref{eq:phione} into \eqref{eq:brak1} gives us
\begin{align}
    \mathcal{F}^*\theta_{u'}(X_u) &= \langle (\Phi \cdot g)^{-1} \cdot e_a \Phi_b^a dx^b , \bar{X}_x \rangle \label{eq:brak2}
\end{align}
Evaluating $(\Phi \cdot g)^{-1} \cdot e_a$

\begin{align}
    (\Phi \cdot g)^{-1} \cdot e_a &= (g^{-1} \cdot \Phi^{-1}) \cdot e_a \\
    &= (g^{-1} \cdot \Phi^{-1})_a^c e_c \\
    &= (g^{-1})_d^c (\Phi^{-1})_a^d e_c \label{eq:dot1}
\end{align}
Substituting equation \eqref{eq:dot1} into \eqref{eq:brak2} gives us
\begin{align}
\mathcal{F}^*\theta_{u'}(X_u) &= \langle (g^{-1})_d^c (\Phi^{-1})_a^d  \Phi_b^a dx^b e_c, \bar{X}_x \rangle  \\
&= \langle (g^{-1})_d^c \delta_b^d dx^b e_c, \bar{X}_x \rangle \\
&= \langle (g^{-1})_b^c  dx^b e_c, \bar{X}_x \rangle \\
&= k_u (\pi_* X_u) \\
\mathcal{F}^*\theta_{u'}(X_u) &= \theta_u (X_u)\qed 
\end{align}
Hence, equation \eqref{eq:ditheta} has been proven. The demonstrated Diffeomorphism Invariance of the Canonical 1-form implies that a theory with the Canonical 1-form as its dynamical variable, such as TEGR, will be a theory with a Diffeomorphism Invariance so long as the other structures in the theory also possess Diffeomorphism Invariance.

\section{Diffeomorphism Invariance of the Flatness of the Connection \label{chap:condifinv}}

In this section, the Diffeomorphism Invariance of the other fundamental structure of TEGR, and by extension AGT, namely its flat Frame Bundle Connection will be examined. 

Given a flat Frame Bundle connection as specified earlier in equation \eqref{eq:phidphi}, 
\begin{equation}
    A = \phi^{-1} d\phi
\end{equation}

Under a pull-back along some diffeomorphism, such a connection transforms according to  
\begin{align}
    A' &= f^* A \\
    &= f^* [\phi^{-1}d\phi] \label{eq:fa} 
\end{align}
Since $\phi$ and $\phi^{-1}$ are 0-forms, we can evaluate equation \eqref{eq:fa} in the following way
\begin{align}
    A' &= f^* [\phi^{-1} \wedge d\phi] \\
    &= f^* \phi^{-1} \wedge f^* d\phi \\
    &= \phi^{-1} \cdot f \wedge d f^* \phi \\
    &= \phi^{-1} \cdot f \wedge d \phi \cdot f 
\end{align}
which we can rewrite as 
\begin{equation}
    A' = \phi'^{-1}d\phi' 
\end{equation}
where
\begin{align}
    \phi'^{-1} &= \phi^{-1} \cdot f \\
    \phi' &= \phi \cdot f
\end{align}

As shown before in equation \eqref{eq:flat}, that $F=0$ for $A=\phi^{-1}d\phi$, it follows that $F'=0$ also for $A'=\phi'^{-1}d\phi'$. This shows that the pull-back of this flat connection along some diffeomorphism yields another connection of the same form. As such, the flatness of the Frame Bundle connection is preserved under an arbitrary diffeomorphism. In other words, the flatness of the Frame Bundle Connection is also diffeomorphism invariant, just as the canonical 1-form is diffeomorphism invariant. Furthermore, since this flat connection is not dynamical and therefore can be freely chosen, as long as the connection is a flat connection of the form $A=\phi^{-1}d\phi$, we conclude that diffeomorphism has no physical implication relating to the effects of diffeomorphism on the Frame Bundle connection. 

\section{Diffeomorphism Invariance of Theories\label{chap:difinvtheories}}

In the previous two sections, it has been established that the two fundamental structures of TEGR, the flatness of the Frame Bundle connection and the canonical 1-form, possess Diffeomorphism Invariance. However, such deductions are merely mathematical, and the implications of these mathematical facts on physical theories, TEGR and AGT, needs to be further laid out. This section will address the question of Diffeomorphism Invariance at the level of these two theories. 

The Diffeomorphism Invariance of a theory is formally defined in the following way. A particular model in that theory which consists of fields of the physical content of the universe (e.g. matter), $\Psi_i$, and relevant dynamical fields, $P_i$, is denoted by
\begin{equation*}
    M(\Psi_i, P_i)
\end{equation*}
Under a diffeomorphism, $d$, the given model transforms into
\begin{equation*}
    M'(d*\Psi_i, d*P_i)
\end{equation*}
where $*$ denotes the push-forwards or the pull-backs according to the type of the relevant fields. 
A theory possesses Diffeomorphism Invariance if $M'$ remains a model of the theory given that $M$ is a model of the theory, for all diffeomorphisms, $d$.

\subsection{Diffeomorphism Invariance in TEGR\label{sec:difinvtegr}}

We recall that in TEGR, as put forward by Aldrovandi and Pereira in \cite{Pereira1Aldrovandi}, the relevant field of TEGR is torsion, which depends on both the flatness of the Frame Bundle connection and the dynamical canonical 1-form. 

In section \ref{chap:candifinv} it has been proven that the canonical 1-form is invariant under arbitrary diffeomorphisms, while in section \ref{chap:condifinv} it has been shown that a flat Frame Bundle Connection remains flat under arbitrary diffeomorphisms. Since torsion only depends on these two facts, we deduce that torsion itself is diffeomorphism invariant, and the corresponding Lagrangians that are constructed from Torsion will be invariant. Thus, TEGR is Diffeomorphism Invariant in the formal sense of: 
\begin{equation*}
    M'(d*\Psi_i, d^*\omega, d^*\theta, ...)
\end{equation*}
is a model of TEGR if
\begin{equation*}
    M(\Psi_i, \omega, \theta, ...)
\end{equation*}
is a model of TEGR, for all diffeomorphisms, $d$, since $d^*\omega$ remains flat if $\omega$ is flat, and $d^*\theta=\theta$.

We can make a stronger claim than the simple fact of the Diffeomorphism Invariance of TEGR, any formulation of TEGR, or any theory with those two structures, or torsion as the dynamical variable are necessarily diffeomorphism invariant. This is provided that no other structure that is not Diffeomorphism Invariant is admitted into the theory in that unspecified $...$ in the models of TEGR. This caveat is not trivial.

\subsection{Diffeomorphism Invariance in AGT\label{sec:difinvagt}}

The main subject matter of this paper, the Diffeomorphism Invariance of AGT defined on the Affine Bundle, remains to be demonstrated. Diffeomorphism Invariance have only been shown in the Frame Bundle. After all, the canonical 1-form and Frame Bundle connection are defined on the Frame Bundle. The AGT, however, is formulated in the closely, but not trivially, related Affine Bundle. Of course, the Affine connection is locally given by equation \eqref{eq:ft}, i.e. $\tilde{F} = T$, which simply means that it is equivalent to the local representation of torsion as defined on the Frame Bundle. As such, it can already be concluded from section \ref{sec:difinvtegr} that since Physics is done in the local representation, AGT which is locally equivalent to TEGR has Diffeomorphism Invariance. 

However, more can be said if we examine AGT at the Principal Bundle level. We recall that earlier in section \ref{chap:ans}, equation \eqref{eq:ft} is obtained through the privileged section defined in \eqref{eq:priv} which reads, 
\begin{equation}
    \tilde{\sigma} = \gamma \cdot \sigma
\end{equation}

This privileged section is composed of a section, $\sigma$, in $LM$ and the special $\gamma$ map which realizes a choice of the origin of the affine frames. With this $\gamma$ map, we establish the correspondence between the Affine connection and the Frame Bundle connection and the canonical 1-form given in equation \eqref{eq:reltilnotil} which reads 
\begin{equation}
    \gamma^*\tilde{\omega} = \omega + \theta
\end{equation}
Under a diffeomorphism, this becomes
\begin{equation}
    \gamma^*\tilde{\omega'} = \omega' + \theta'
\end{equation}
And since the canonical 1-form is diffeomorphism invariant, we get
\begin{equation}
    \gamma^*\tilde{\omega'} = \omega' + \theta
\end{equation}
Furthermore, the flatness and form of $\omega$ is preserved rendering $\omega'$ physically equivalent to $\omega$. Therefore, Diffeomorphism Invariance in AGT, at the Principal Bundle level, means that  $\gamma^*\tilde{\omega'}$ is physically equivalent to $\gamma^*\tilde{\omega}$ . Thus, given a choice of origins of the Affine frames, realized by $\gamma$, AGT possesses Diffeomorphism Invariance in the formal sense of: 
\begin{equation*}
    M'(d*\Psi_i, d^*\gamma^*\tilde{\omega'}, ...)
\end{equation*}
is a model of AGT if
\begin{equation*}
    M(\Psi_i, \gamma^*\tilde{\omega}, ...)
\end{equation*}
is a model of AGT, for all diffeomorphisms, $d$.

\section{Translational Gauge Invariance and Diffeomorphism Invariance\label{chap:relationship}}

The relationship between Translational Gauge Invariance and Diffeomorphism Invariance in the context of theories of gravity defined on the Affine Bundle has long been a matter of contention among physicists. Gronwald in his highly cited paper, \cite{Gronwald}, laid out an argument regarding the relationship between Translational Gauge Invariance and Diffeomorphism Invariance. He argues that given a choice of origins in the Affine frames of an Affine Bundle, a soldering is performed. As a result of this soldering, the Affine Bundle is reduced to the Frame Bundle, and Translational Gauge freedom is lost. This choice of origins is nothing but the $\gamma$ that we established in section \ref{chap:ans} and discussed in the previous section.  However, according to Gronwald, the soldering allows one to establish a one-to-one correspondence between a general gauge transformation and diffeomorphisms, via the idea of the development of curves. The idea of the development of curves can be found in \cite{Kobayashi} section  III.4. The lost of translational gauge is then regained through the corresponding diffeomorphisms. 

While Gronwald's description of $\gamma$ is eye-opening, his arguments of the correspondence between Translational Gauge Invariance and Diffeomorphism Invariance is less convincing. What Gronwald has established is merely a one-to-one correspondence between a translation gauge transformation of the Affine Bundle and a diffeomorphism of the base manifold, not between their symmetries. The question of their actual symmetries and the relationship between the symmetries remain unanswered. 

After the rigorous discussion and proofs of both the Translational Gauge Invariance and Diffeomorphism Invariance of AGT that has been given in this paper, we are now in a good position to examine the relationship between the two symmetries. 

In section \ref{chap:transinv}, we have shown that AGT possesses gauge invariance related to its translational gauge degree of freedom. Equations \eqref{eq:one}, \eqref{eq:two}, and \eqref{eq:three} clearly shows an explicit translational gauge invariance. On the other hand, as has been clearly demonstrated in section \ref{sec:difinvagt}, the pull-back of the Affine connection along the $\gamma$ map, i.e. $\gamma^*\tilde{\omega}$ , has Diffeomorphism Invariance. That is, the Affine connection, together with a soldering, renders that connection, and its local representations, Diffeomorphism Invariant. We explicitly see how the soldering, realized by $\gamma$ breaks the translational gauge freedom as it picks out a privileged section $\tilde{\sigma}$, while at the same time endowing AGT with a Diffeomorphism Invariance. 

In conclusion, in the AGT, without the soldering, we have Translational Gauge Invariance, and with the soldering, we have Diffeomorphism Invariance and no Translational Gauge freedom. This is how the two symmetries relate and this thesis have shown it rigorously.

\nocite{*}
\bibliography{aipsamp}

\end{document}